\documentclass[12pt,reqno,centertags,a4paper]{iopart}
\expandafter\let\csname equation*\endcsname\relax
\expandafter\let\csname endequation*\endcsname\relax

\usepackage[T1]{fontenc}
\usepackage{
amsfonts,
amssymb,
amsthm
}

\usepackage{pslatex}
\usepackage{enumerate}
\usepackage{nicefrac,xfrac}
\usepackage{latexsym}
\usepackage[classicReIm,uprightRoman,upright,frenchstyle,
uprightgreeks,notextcomp]{kpfonts}
\usepackage{cite}

\allowdisplaybreaks
\numberwithin{equation}{section}

\newcommand{\R}{\mathbb{R}}
\newcommand{\C}{\mathbb{C}}
\newcommand{\Dw}{\mathscr{D}}

\newcommand{\X}{\mathscr{X}}

\newcommand{\oll}[1]{\overline{#1}}
\newcommand{\ot}{\text{\large $\otimes$}}
\newcommand{\abs}[1]{\lvert#1\rvert}

\newcommand{\matr}[2]{\langle #1,#2\rangle}

\DeclareMathOperator{\Ima}{Im}
\DeclareMathOperator{\Ker}{Ker}

\begin{document}

\title{A note on the spectrum of a two-particle Rashba Hamiltonian}

\author{ R. Jur\v{s}\.{e}nas }
 
\address{ Institute of Theoretical Physics and Astronomy of Vilnius University, \\
A.~Go\v{s}tauto 12, 01108 Vilnius, Lithuania }

\begin{abstract}
In a series of recent papers it was shown that, when the attractive $s$-wave interaction is dominant, 
the spin-orbit coupled fermions form a bound state. Attributed to a convenient momentum representation,
it became a common condition of agreement to express the bound state as a function of the
center-of-mass momentum $Q$. In this letter we prove that the bound state of Rashba fermions does
not depend on the chosen representation. That is, all the states characterized by nonzero $Q$ fail 
to obey the translation symmetry.
\end{abstract}


{\small
\tableofcontents
}

\maketitle

\section{Introduction}\label{sec:s1}

The role of a bound state is in the spotlight of neutral Fermi atoms.
The motives of research, technical possibilities and
important physical issues are discussed in many 
papers \cite{Vyasanakere11,Yu11,Iskin11,Jiang11,Cui12,Takei12,Dong13}
where---among other celebrated results---the existence of a bound state is 
determined by establishing the poles of the Fourier transform of Green's function.
Also \cite{Takei12}, one can explore a free two-particle Green's function in
deriving the so-called self-consistency condition which is just the weak 
(or distributional) solution to the Schr\"{o}dinger equation. Although trivial 
in many applications, the pole itself is insufficient to determine the bound state. 
According to the Aronszajn--Donoghue spectral theory of rank one perturbations
\cite{Albeverio00,Simon05} [$s$-wave interaction is exactly rank one],
the derivative of the free Green's function with respect to (w.r.t.)
the eigenvalue must be finite. Otherwise, the spectral points are in the
singular continuous spectrum. This is exactly the case when one says
\cite{Takei12} that the bound state ceases to exist.

The single-particle Rashba Hamiltonian in $L_{2}(\R^{2};\C^{2})=:\X$, 
of a particle with mass $m$ and Rashba spin-orbit coupling $\alpha$ (we use the 
$\hbar=c=1$ units) is invariant under Euclidean moves. If parametrized in
terms of the center-of-mass $\vec{R}\in\R^{2}$ and the relative $\vec{r}\in\R^{2}$
coordinates, the same applies to the Hamiltonian in 
$(\X,d^{2}\vec{R})\ot(\X,d^{2}\vec{r})$ which describes the
two interacting particles. For the dominant $s$-wave interaction, with the 
interaction strength $\gamma$, the Euclidean group is naturally reduced to the
translation group. The latter is represented as a tensor product of two
two-dimensional groups whose irreducible representations act on
$\vec{R}$ and $\vec{r}$, respectively. Due to the interaction, it is clear
that the two-particle Hamiltonian commutes with the generator $\vec{P}\ot I$,
where $i\vec{P}$ is the gradient in $\vec{R}$ and $I$ is the identity operator 
in $(\X,d^{2}\vec{r})$, whereas it does not commute with the generator in
$\vec{r}$. The commutation relation indicates that the eigenspace of the
Hamiltonian is a dense subset in the domain of the closure of the
generator in the Sobolev norm (apply eg \cite[Corollary~A.7.3.6]{Bhattach12}). 
Subsequently, the generalized eigenvectors of the Hamiltonian are labeled by its 
spectral points $\lambda$ along with the spectral points $\vec{\Lambda}\in\R^{2}$ 
of the closure of $\vec{P}$. 

In this letter we shall prove that only the translation invariant eigenvectors,
that is those with $\vec{\Lambda}=\vec{0}$, are the eigenvectors of the
two-particle Rashba Hamiltonian with point-interaction. For other
two-particle interactions, this is not necessarily the case though
(see discussion in Sec.~\ref{sec:s6}).

\section{Translation symmetry}\label{sec:s2}

Consider an arbitrary $f=f(\vec{R},\vec{r})$ in the Schwartz space 
$\Dw(\R^{4};\C^{4})$ of smooth functions with compact support.
Then $f$ has a Fourier transform $\chi=\chi(\vec{Q},\vec{k})$, where 
$\vec{Q}\in\R^{2}$ denotes the center-of-mass momentum and $\vec{k}\in\R^{2}$ 
is the relative momentum.

Since $\Dw(\R^{4};\C^{4})$ is a dense subspace of the Hilbert space
$L_{2}(\R^{4};\C^{4})$, denoted $\X_{o}$, and $\X\ot\X$ is isomorphic to $\X_{o}$,
one can restate the irreducible representation $\exp(i\vec{P}\cdot\vec{a})\ot I$ 
(all $\vec{a}\in\R^{2}$) of the translation group by simply writing $\exp(i\vec{P}\cdot\vec{a})$. 
Suppose that $\vec{\Lambda}\in\R^{2}$ and that $f$ solves 
$(\vec{P}-\vec{\Lambda})f=\vec{0}$. Then $f$ is of the form

\begin{equation}
f(\vec{R},\vec{r})=e^{i(\vec{\Lambda}\cdot\vec{R})}\varphi(\vec{r}),\quad 
\varphi\in \Dw(\R^{2};\C^{4}).
\label{eq:f}
\end{equation}

\noindent{}Then the Fourier transform reads

\begin{equation}
\chi(\vec{Q},\vec{k})=(2\pi)^{2}\delta(\vec{Q}-\vec{\Lambda})\hat{\varphi}(\vec{k}),
\label{eq:chi}
\end{equation}

\noindent{}where $\hat{\varphi}\in \Dw(\R^{2};\C^{4})$ is the Fourier transform of $\varphi$ and
$\delta(\cdot)$ is a two-dimensional Dirac distribution. It appears from (\ref{eq:chi}) that
the Fourier transform $\chi$ represents a singular
distribution, the existence of which is predetermined by the relation $\vec{Q}=\vec{\Lambda}$.
In particular, (\ref{eq:f}) implies that for $\vec{\Lambda}=\vec{0}$,
$f$ is translation invariant. In general, it also follows from (\ref{eq:f}) that $f$ is labeled by
the spectral point $\vec{\Lambda}$, while $\varphi$ is independent of $\vec{\Lambda}$.

\section{Green's function}\label{sec:s3}

The single-particle Rashba Hamiltonian in $\X$ is realized through the differential
expression $-(2m)^{-1}\Delta+\alpha U$, where $U:=-i(\nabla_{y}\sigma_{x}-\nabla_{x}\sigma_{y})$.
Based on the relation $U^{2}=-\Delta$, the associated Green's function can be obtained in
an extremely elegant way \cite{Bruning07-2}. In coordinate representation, it 
is a linear combination of the Bessel functions of the second kind. It can be shown \cite{Jursenas13}
that the single-particle Green's function is sufficient to recover the spectrum of the
two-particle Rashba Hamiltonian with point-interaction. The bound state $\lambda$ corresponds to
the Fourier transform of the Hamiltonian with $\vec{Q}=\vec{0}$. More precisely, given
$m=1/2$, $\alpha>0$, $\gamma<0$, a single bound state is found to be equal to $\lambda=-\alpha^{2}/(2\sin^{2}\omega)$,
where the parameter $-\pi/2<\omega<0$ solves the transcendental equation
$j_{\gamma}+\ln(\alpha/4)=\ln\abs{\sin\omega}-\omega\abs{\tan\omega}$, where one defines
$j_{\gamma}:=4\pi/\abs{\gamma}-\Psi(1)$, and $\Psi(1)$ is the digamma function. The result
is in exact agreement with the associated self-consistency condition (see the integral in \cite[eq.~(31)]{Takei12})
obtained by considering the free two-particle Green's function in momentum representation. Moreover, the above
equation allows one to evaluate the characteristic radius of the interaction potential which is found to be
$\exp(-\Psi(1))/2\simeq0.89$ (all $\alpha>0$, all $\gamma<0$).

Let $g_{\mu}^{0}(\vec{Q},\vec{k})$ be the free two-particle Green's function in momentum representation projected onto
the singlet basis \cite{Takei12}; $\mu\in\C\backslash[-\alpha^{2}m,\infty)$. Then $g_{\mu}^{0}$ is a bounded everywhere
defined function on $\R^{4}$. The function $g_{\mu}(\vec{Q},\vec{k})$ corresponding to the projected integral kernel of
the two-particle Hamiltonian with point-interaction fulfills the resolvent identity

\begin{align}
g_{\mu}(\vec{Q},\vec{k})=&
g_{\mu}^{0}(\vec{Q},\vec{k})\left(1-\gamma \tilde{g}_{\mu}(Q) \right), \quad\text{where} \nonumber \\
g_{\mu}^{0}(\vec{Q},\vec{k}):=&(\epsilon-\mu)((\epsilon-\mu)^{2}-\alpha^{2}Q^{2})\bigl(
(\epsilon-\mu)^{4} \nonumber \\
&-4\alpha^{2}m\epsilon(\epsilon-\mu)^{2}+4\alpha^{4}(\vec{Q}\cdot\vec{k})^{2} \bigr)^{-1},
\nonumber \\
\epsilon(Q,k):=&\frac{k^{2}}{m}+\frac{Q^{2}}{4m}, \nonumber \\
\tilde{g}_{\mu}(Q):=&\int_{k\leq C}\frac{d^{2}\vec{k}}{(2\pi)^{2}}\;
g_{\mu}(\vec{Q},\vec{k}).
\label{eq:Green}
\end{align} 

\noindent{}The condition $k\leq C$, where $C$ is the UV cutoff, ensures that the integral exists and
the obtained natural logarithm is finite.

It appears from the resolvent identity that the (projected) Green's function

\begin{equation}
g_{\mu}(\vec{Q},\vec{k})=\frac{g_{\mu}^{0}(\vec{Q},\vec{k})}{1+\gamma\tilde{g}_{\mu}^{0}(Q)}
\label{eq:GreenV}
\end{equation} 

\noindent{}where $\tilde{g}_{\mu}^{0}$ is defined similar to $\tilde{g}_{\mu}$ but with $g_{\mu}$ replaced by $g_{\mu}^{0}$.
Then the denominator of the $g_{\mu}$ determines the bound state solution unless 
$\abs{(\partial/\partial\mu)\tilde{g}_{\mu}^{0}}$ is sufficiently large
at $\mu=\lambda$ (where $\lambda$ solves $1+\gamma\tilde{g}_{\lambda}^{0}=0$). We use the notion <<sufficiently
large>> rather than <<infinite>> because one accounts for $k\leq C$ but not for $k<\infty$ in the integral in
(\ref{eq:Green}). By (\ref{eq:GreenV}), the bound state solution is a function of the center-of-mass
momentum, $\lambda=\lambda(Q)$. The solution agrees with that obtained from
the single-particle Green's function only if $Q=0$. By using a more general Aronszajn--Donoghue spectral theory,
it will be shown hereafter that the values of $Q$ other than zero do not represent
bound state solutions.

\section{Nevanlinna function}\label{sec:s4}

In virtue of (\ref{eq:Green}), equation~(\ref{eq:GreenV}) points to 
$\tilde{g}_{\mu}=\tilde{g}_{\mu}^{0}/(1+\gamma\tilde{g}_{\mu}^{0})$. In turn, the expression
for $\tilde{g}_{\mu}$ alludes to a well-known Aronszajn--Krein formula for the symmetric rank one bounded
perturbation of a self-adjoint operator in the Hilbert space. Without referring to the extension theory of
symmetric operators, one can show \cite[Theorem~1.1.1]{Albeverio00} that, if given a self-adjoint
operator $H$, then the subtraction of the resolvents of the perturbed operator 
$H_{\gamma}:=H+\gamma\matr{\delta}{\cdot}\delta$ and that of $H$ is proportional to
$1/(1+\gamma F_{\mu})$, where $F_{\mu}:=\matr{\delta}{(H-\mu)^{-1}\delta}$ is known as the
Borel transform of a measure, and it belongs to the Nevanlinna class (that is,
the complex conjugate $\oll{F_{\mu}}=F_{\oll{\mu}}$ and the imaginary part 
$\Ima F_{\mu}/\Ima\mu\geq0$, provided $\mu$ is in the resolvent set of $H_{\gamma}$). Here
$\delta$ is the Dirac distribution and $\matr{\cdot}{\cdot}$ denotes the scalar product in
a concrete Hilbert space.

Let us study the function $F_{\mu}$, applied to our case, in a more detailed fashion.
For this, let $H$ be a free two-particle Rashba Hamiltonian in the Hilbert space $\X_{o}$.
$H$ is parametrized in terms of the center-of-mass $\vec{R}$ and the relative $\vec{r}$
coordinates. Then the $s$-wave interaction can be estimated using the Dirac distribution $\delta$
at the relative coordinate $\vec{r}\in\R^{2}$ (the interaction strength is $\gamma$).
As a result, the form sum $H_{\gamma}$ describes the two-particle system with
point-interaction. To see this explicitly, it suffices to note that for $f=f(\vec{R},\vec{r})$
in the domain of $H$, it holds $\matr{\delta}{f}\delta=f(\vec{R},\vec{0})\delta=f\delta$.

The Aronszajn--Krein formula reads \cite[Eq.~(11.13)]{Simon05}

\begin{equation}
(H_{\gamma}-\mu)^{-1}=(1+\gamma F_{\mu})^{-1}(H-\mu)^{-1}.
\label{eq:Krein}
\end{equation}

\noindent{}It takes little effort to verify that equation~(\ref{eq:Krein})
eventually leads to the expression of the Green's function $g_{\mu}$ in 
(\ref{eq:GreenV}) but with $\tilde{g}_{\mu}^{0}$ replaced by $\tilde{F}_{\mu}$,
where $\tilde{F}_{\mu}$ denotes the Nevanlinna function $F_{\mu}$ projected onto the singlet basis.
By inspection, the Fourier transform

\begin{equation}
\widehat{((H-\mu)^{-1}\delta)}(\vec{Q},\vec{k})=(2\pi)^{2}\delta(\vec{Q})
\hat{G}_{\mu}^{0}(\vec{0},\vec{k}),
\label{eq:Fourier}
\end{equation}

\noindent{}where $\hat{G}_{\mu}^{0}$ is the Green's function of $H$
in momentum representation. The projection of the Green's function onto 
the singlet basis is given by $g_{\mu}^{0}$. By (\ref{eq:Fourier})
and the fact that the norm of arbitrary $f\in\X_{o}$ coincides with the norm of 
its Fourier transform, one easily deduces that the Nevanlinna function 

\begin{equation}
\tilde{F}_{\mu}=\tilde{g}_{\mu}^{0}(0).
\label{eq:Nevan}
\end{equation}

\noindent{}Bringing together (\ref{eq:GreenV}) and (\ref{eq:Nevan}), one figures out that
the zero-range interaction $\delta$ admits only the bound state that corresponds to $Q=0$, while the
free Green's function $g_{\mu}^{0}$ (recall the definition in (\ref{eq:Green}))
accepts all possible values $\vec{Q}\in\R^{2}$. 
The affirmation of the result for other types of spin-orbit coupling comes from the 
Nevanlinna function, for the reason that the Green's function was not specified
when deriving (\ref{eq:Nevan}).

The obtained result, (\ref{eq:Nevan}), is not accidental, as it is closely related to the
eigenvectors obeying the translation symmetry.

\section{Eigenspace}\label{sec:s5}

The fact that the bound state of two interacting Rashba particles exists only for
the zero-valued center-of-mass momentum can be affirmed by the commutation property
of the Hamiltonian $H_{\gamma}$ with the generator $\vec{P}$ of the translation group.
The commutation relation designates the spectral points $\vec{\Lambda}\in\R^{2}$
of $\vec{P}$ to the eigenspace of $H_{\gamma}$. As a result, the eigenvectors of
the Hamiltonian are of the form (\ref{eq:f}). Since the Hamiltonian is defined as
the operator in the Hilbert space $\X_{o}$, it should be emphasized that in this case,
equation~(\ref{eq:f}) (as well as (\ref{eq:chi})) is understood in the generalized sense;
by the density result \cite{Bhattach12}, the Schwartz space can be extended to the
appropriate domain of definition of $H_{\gamma}$.

Let $\Omega$ be the projection onto the singlet basis. In virtue of 
(\ref{eq:f})--(\ref{eq:chi}), vectors $f$ in the eigenspace  $\Ker(H_{\gamma}-\lambda)$ 
fulfill the eigenvalue equation

\begin{equation}
\hat{\varphi}(\vec{k})+\gamma\hat{G}_{\lambda}^{0}(\vec{\Lambda},\vec{k})
\int_{\R^{2}}\frac{d^{2}\vec{k}^{\prime}}{(2\pi)^{2}}\Omega \hat{\varphi}(\vec{k}^{\prime})=0.
\label{eq:eigenvalue}
\end{equation}

\noindent{}By the fact that $\varphi$ is independent of $\vec{\Lambda}$, equation~(\ref{eq:eigenvalue})
yields $(\partial/\partial\vec{\Lambda})\hat{G}_{\lambda}^{0}=\vec{0}$. That is, 
$\hat{G}_{\lambda}^{0}(\vec{\Lambda},\vec{k})$ is a constant w.r.t. $\vec{\Lambda}\in\R^{2}$.
The only one nontrivial solution to the latter equation w.r.t. $\vec{\Lambda}\in\R^{2}$ reads
$\Lambda=0$. By (\ref{eq:f}), this implies that $f=\varphi$ or else, the space of translation
invariant eigenvectors is reduced from $L_{2}(\R^{4};\C^{4})$ to $L_{2}(\R^{2};\C^{4})$. The
cancellation of one spatial vector becomes clear recalling that $\X_{o}$ is isomorphic to the
tensor product of spaces $(\X,d^{2}\vec{R})\ot(\X,d^{2}\vec{r})$. The Hamiltonian in the Hilbert space
$(\X,d^{2}\vec{R})$ is parametrized in terms of the center-of-mass coordinate $\vec{R}$ and it does not 
have bound states, while the Hamiltonian in the space $(\X,d^{2}\vec{r})$ is parametrized in terms of the 
relative coordinate $\vec{r}$ and it has a bound state due to the Dirac distribution at $\vec{r}$. This latter
space is exactly the one where the eigenspace of the two-particle Rashba Hamiltonian is situated.

\section{Conclusion and discussion}\label{sec:s6}

An attempt to clarify the issue regarding the bound state of spin-orbit coupled
fermions was the primary intent of this note. 
The results of this letter clearly demonstrate that the eigenvector of the
two-particle Rashba Hamiltonian with point-interaction $\delta$ is translation invariant.
The associated bound state exists only for the zero-valued center-of-mass momentum $Q$.

The main conclusion of the present report is
that, if the two-particle Hamiltonian $H_{\gamma}$, where $\gamma$ is the strength of
interaction, possesses translation symmetry, then for any 
two-particle interaction of rank one, the bound state exists only for the zero-valued
center-of-mass momentum $Q$. The same applies to other types of spin-orbit coupling.
On the other hand, if the Hamiltonian is not translation invariant, the
conclusion does not necessarily hold. To explain this, let us give some
examples.

The rank one perturbation to the free
Hamiltonian $H$ is characterized by the term $\gamma\matr{\phi}{\cdot}\phi$, where
$\phi$ need not be the unit vector in $\X_{o}$; it even need not be in the Hilbert space
\cite{Albeverio00,Simon05,Schmudgen12}. In this case the Nevanlinna function
$F_{\mu}$ reads $\matr{\phi}{(H-\mu)^{-1}\phi}$. For the proper potential (that is, the
vector $\phi$), one can possibly derive bound states with nonzero $Q$. For example,
let $\phi$ be summable on rectangle $\R^{2}\times\R^{2}$. Then an easy calculation
shows that

\begin{equation}
\widehat{((H-\mu)^{-1}\phi)}(\vec{Q},\vec{k})=
\hat{\phi}(\vec{Q},\vec{k})\hat{G}_{\mu}^{0}(\vec{Q},\vec{k}),
\label{eq:Fourier2}
\end{equation}

\noindent{}where $\hat{\phi}$ is the Fourier transform of $\phi=
\phi(\vec{R},\vec{r})$. It is clear from (\ref{eq:Fourier2})
that whenever $\phi(\vec{R},\vec{r})=\phi(\vec{r})$, that is,
when the two-particle Hamiltonian is invariant under translations,
it always holds

\begin{equation}
\widehat{((H-\mu)^{-1}\phi)}(\vec{Q},\vec{k})=
(2\pi)^{2}\delta(\vec{Q})
\hat{\phi}(\vec{k})\hat{G}_{\mu}^{0}(\vec{0},\vec{k})
\label{eq:Fourier3}
\end{equation}

\noindent{}hence $Q=0$, no matter what type of interaction between two particles
is specified, viz. zero-range or short-range. In particular,
choosing $\phi=\delta$, (\ref{eq:Fourier3}) formally coincides with
(\ref{eq:Fourier}). On the other hand, if, for example, given the vector
$\phi(\vec{R},\vec{r})=\exp(i(\vec{a}\cdot\vec{R}))\delta(\vec{r})$ for some $\vec{a}\in\R^{2}$, 
one obtains from (\ref{eq:Fourier2}) that $\vec{Q}=\vec{a}$, and thus the
bound state depends on the controllable parameter $a$; in this case
the two-particle Hamiltonian is not translation invariant and
the (projected) Nevanlinna function $\tilde{F}_{\mu}=\tilde{g}_{\mu}^{0}(a)$.

\subsection*{Acknowledgments}

The work was supported by the Research Council of Lithuania (No.~VP1-3.1-\v{S}MM-01-V-02-004).
The author thanks B. M. Anderson for enlightening him with certain aspects related to the 
self-consistency condition \cite{Takei12}. It is a pleasure to thank G. Juzeli\={u}nas
for discussions. The author is very grateful to anonymous referee for his
careful reading of the manuscript and the remarks which helped to improve it.

\bigskip{}
\bigskip{}
\bigskip{}

\bibliographystyle{apsrev4-1}

\providecommand{\noopsort}[1]{}\providecommand{\singleletter}[1]{#1}%

\end{document}